\RequirePackage{fix-cm}
\documentclass{svjour3}                     
\usepackage{subfigure}
\usepackage{xcolor}
\usepackage[colorlinks=true,linkcolor=red,citecolor=blue,urlcolor=cyan]{hyperref}
\usepackage{mathtools}
\usepackage{fullpage}
\usepackage{float}
\usepackage{xcolor, soul}
\sethlcolor{green}
\usepackage{amsmath}
\usepackage{hyperref}
\usepackage{amssymb}
\usepackage{mathtools}
\usepackage{xfrac}
\usepackage{breqn}
\usepackage{amsfonts}
\usepackage{graphicx,epstopdf}
\begin{document}
	\title{Multi-lump, lump-kink and interaction of breather with other nonlinear waves of a couple Boussinesq system}
	\author{Snehalata Nasipuri \and Prasanta Chatterjee*}	
	\institute{Snehalata Nasipuri \at Department of Mathematics, Visva-Bharati, Santiniketan-731235, India\\
		\email\href{mailto: snehalatanasipuri@gmail.com}{snehalatanasipuri@gmail.com}     
		\and
		Prasanta Chatterjee* \at
		Department of Mathematics, Visva-Bharati, Santiniketan-731235, India\\
		\email\href{mailto:prasantachatterjee1@ iffmail.com}{prasantachatterjee1@rediffmail.com}
	}
	\date{Received: date / Accepted: date} 
	\maketitle
	\begin{abstract}
		We investigate the interaction characteristics of nonlinear coherent structures in the couple Boussinesq (CB) system using the Hirota bilinear approach. First, we derive the lump solutions using a positive quadratic polynomial within the Hirota perturbation technique. Next, We study the explicit interactions between lumps and one or two-kink waves, and observe double lump patterns. Furthermore, we discover the interactions of breathers, revealing a diffusion-like behavior. We notice that breather waves can interact with periodic, kink, and bright solitons for specific parameter sets in the CB system. Interactions between two chains of double breathers are also found. We analyze our results using a combination of symbolic computations and graphical representations, providing a deeper understanding of their behavior. This study reveals previously unreported nonlinear dynamics in the CB system.
		
	\end{abstract}
	\section{Introduction}
	It is well-established that solitary wave solutions of nonlinear partial differential equations (NPDEs) can provide valuable insights into various physical and biological phenomena \cite{in1.1}-\cite{mkdv-sin}. As a result, the search for solitary wave solutions of NPDEs has become an active area of research. Over the years, numerous methods have been developed to find these solutions, including the inverse scattering method \cite{IST1},\cite{IST2}, Darboux transformation method \cite{DT1}, Hirota's bilinear method \cite{hirota1}-\cite{hirota3}, Painlev\'{e} expansion method \cite{pain}, Homogeneous balance method \cite{hbalance},  Jacobian elliptic function expansion method \cite{jacobian}, etc. In recent decades, researchers have built several coupled and multi-component nonlinear evolution equations to describe complex nonlinear phenomena across various branches of physics, encompassing fluid dynamics, optics, plasma physics, and solid-state physics \cite{1}-\cite{5}, etc. For instance, the interaction between long-wave and short-wave packets in nonlinear dispersive media is effectively described by the cubic nonlinear Schr\"{o}dinger-Korteweg-de Vries(NLS-KdV) system \cite{exam}.
	Accurately solving these paired nonlinear equations is crucial for understanding real-world phenomena and has become an important area of research. Obtaining precise solutions to problems involving coupled nonlinear systems enables researchers to better comprehend these phenomena. The investigation of these systems is crucial, as the nonlinear interactions between multiple waves can give rise to new and unexplored features of several nonlinear waves.  To achieve this goal, various methods have been developed, including the homotopy analysis method \cite{homotopy1}, the extended tanh method \cite{extanh1}, the homotopy perturbation method \cite{homo-per1}, the modified extended method \cite{modi-ext1}, the trial equation method \cite{trial-met}, the generalized coupled trial equation method \cite{gen-trial}, the generalized Bernoulli sub-ODE technique \cite{ber-subode}, etc. These methods have significantly advanced our understanding of coupled nonlinear systems. The solutions of nonlinear coupled PDEs such as traveling waves, solitary waves, and solitons (including kink, peakon, bright, breather, anti-kink, dark, periodic, rogue wave, lump, and M-shape solitons), are essential for advancing knowledge in physics, life sciences, and applied sciences. \\
	The Boussinesq-like equations are extensively used in  computational simulations of ocean engineering, particularly for modeling fluids with a pronounced scale disparity, where the vertical scale is significantly smaller than the horizontal scale.
	In 1967 \cite{pereg1}, D.H. Peregrine studied the shallow water wave equations and developed what is now commonly referred to  as the Peregrine Boussinesq system. His work focused on the propagation of long nonlinear waves in shallow water, introducing a Boussinesq-type system that models bidirectional wave motion with higher accuracy by accounting for both nonlinear and dispersive effects. Recent studies have made significant contributions to this field. For example, Ali et al. \cite{cbs3} successfully applied the modified Sardar-sub equation method to obtain soliton solutions and chaotic patterns for the space-time fractional coupled Boussinesq model. Li et al. \cite{cbs4} employed the Ansatz method to derive explicit soliton solutions and complex singular solutions for the  nonlocal Boussinesq equations. Ma et al. \cite{cbs5} have investigated the soliton solutions of the nonlocal Boussinesq equations using the Hirota bilinear method. Ayati et al. \cite{cbs6} have developed the modified simplest method to obtain exact solutions for the coupled generalized Schr\"{o}dinger-Boussinesq system and the Boussinesq-type coupled system. \\\\
	We analyze the coupled Boussinesq system, given in the form 
	\begin{equation}\label{m1}
		\begin{split}
			u_t +v_x +2uu_x &=0, \\ v_t +2(uv)_x +u_{xxx}&=0.
		\end{split}
	\end{equation}
	This system arises in the study of fluid flow, specifically describing the propagation of shallow-water waves. In this context, $x$ and $t$ represent the normalized space and time variables, respectively. The function $u(x,t)$ represents the horizontal velocity, specifically the depth-averaged horizontal velocity field at the leading order. In contrast,  $v(x,t)$ describes the height of the water surface above the bottom horizontal level.
	These equations are derived from a mechanistic analysis of two-layer shallow water flow, which facilitates the study of wave patterns formed by one fluid overlaying another. Notably, the CB system has been effectively utilized in modeling oil leakage accidents \cite{cb5}.
	Researchers have explored numerous aspects of this system, including its structural properties and potential for exact waveform solutions. Various studies have been conducted on this system, including the work by Ravi et al. \cite{mymodel4}, who employed the Exp-function method to obtain exact solutions. Moreover, Ahmad et al. \cite{mymodel-decomp} utilized the decomposition method, while Liang et al. \cite{mymodel-mapping} determined Jacobian elliptic function-type solutions. A range of other studies have also investigated exact solutions using diverse techniques \cite{mymodel-stability}-\cite{mymodel-homoblnc}, and some have extended their analysis \cite{mymodel-frac1,mymodel2} to the time-fractional CB system. The study on variant Boussinesq equations has produced the solitary wave solutions in \cite{mymodel3,mymodel-g'g}. Additionally, multiple investigations \cite{mymodel-his}-\cite{mymodel-sim-equ} have examined various wave propagations for the CB system \eqref{m1}. Numerical investigations have also been performed by several researchers \cite{mymodel5}-\cite{mymodel-ndm}. \\\\
	Although the aforementioned works highlighted several features of the CB system, still some important applications of lump and breather solutions are yet to be explored. Hence, the primary objective of this investigation is to study the multi-lump, lump-kink, and  multi-breather interactions of the CB system \eqref{m1}. To achieve this, we employ the HBM and conduct a comprehensive analysis of the lump and breather structures of the CB system. The HBM \cite{h1}, \cite{h3} is an efficient and straightforward approach for studying the interacting properties of various nonlinear features, including lumps, lump-kinks, and breathers. This article is arranged as follows: Sect. \ref{lump}, provides the Hirota bilinear form of the CB system \eqref{m1} using rational transformations and describes the dynamics of single-lump and double-lump solutions. Sect. \ref{lump1kink} and Sect. \ref{lump2kinks} represent the interactions of lumps with one-kink and two-kink waves, respectively. Sect. \ref{bp} discovers the interactions of breathers with other nonlinear waves, including periodic solution, bright soliton, kink wave, and second-order breathers, in the CB system.  Finally, our conclusions and proposals for future research directions are presented in Sect. \ref{con}.

	\section{Lump solutions}\label{lump}
	Lump solitons are fully localized, non-singular solutions that arise in specific nonlinear PDEs, typically on shallow water surfaces where surface tension effects are minimal. Investigating lump solitons is crucial for gaining insight into stable and localized wave phenomena in both space and time. To derive the lump soliton solutions for the system \eqref{m1} using the HBM, at first we take the dependent variable transformations 
	\begin{equation}\label{trn1}
		\begin{split}
			u(x,t)&=\frac{\partial}{\partial x}\left[ln \left( \frac{f}{g} \right) \right], \\ v(x,t)&= \frac{\partial^2}{\partial x^2}\Big[ln \left( fg \right)\Big],
		\end{split}
	\end{equation}
	which reduces the system \eqref{m1} into the following quadratic form
\begin{equation}\label{quadratic}
f_{xt}g-f_tg_x-f_xg_t+fg_{xt}+f_{xxx}g-3f_{xx}g_x+3f_xg_{xx}-fg_{xxx}=0.		
\end{equation}
Now, the Eq. \eqref{quadratic} can be converted to the Hirota bilinear form 
	\begin{equation}\label{hb1}
		\left( D_x D_t + D^3_x \right)(f.g)=0.
	\end{equation}
	Here $D$ denotes the Hirota bilinear operator, defined as follows
	\begin{equation}\label{hb2}
		D_x^p D_t^q (f.g)= \left(\frac{\partial}{\partial x} - \frac{\partial}{\partial x^{\prime}} \right)^p \left(\frac{\partial}{\partial t} - \frac{\partial}{\partial t^{\prime}} \right)^q \Big. f(x,t) g(x^{\prime}, t^{\prime}) \Big \vert_{x^{\prime}=x, t^{\prime}=t}, 
	\end{equation}
	$f(x,t)$ and $g(x,t)$ are any two smooth functions, and $p,q \in \mathbb{N}.$ 
	To demonstrate the spatial localization of lump type waves, we introduce auxiliary functions $f(x,t)$ and $g(x,t)$, expressed as the sum of two positive quadratic functions, as previously established in the literature \cite{mylump}.
	\begin{equation}\label{lump1}
		f(x,t)=(c_0 x+t)^2 +(c_1 x+t)^2 +c_2;~~ g(x,t)=(c_0 x+t)^2 +(c_1 x+t)^2 +c_3, 
	\end{equation}
	where $c_i ~(0 \leq i \leq 3)$ are the specific constants to be determined. Now, substituting expression \eqref{lump1} into Eq. \eqref{quadratic}, a set of algebraic equations is obtained. By accumulating the coefficients of all powers of the variables $x, t$ to zero, we obtain the relation $c_1=-c_0,$ with the free parameters $c_0, c_2, c_3$. Hence the standard lump solution for the system \eqref{m1} is given by 
	\begin{equation}\label{lump2} \begin{split}
			& u(x,t)= - \frac{4 c_0^2 (c_2 -c_3)x}{\left(c_2 + 2(t^2 + c_0^2 x^2)\right) \left(c_3 +2(t^2+c_0^2 x^2) \right)}, \\
			& v(x,t)= \frac{1}{\left(c_2 +2(t^2 + c_0^2 x^2) \right)^2 \left(c_3 +2(t^2 +c_0^2 x^2) \right)^2} \Bigg[4 c_0^2 \Bigg(c_2^2 \left(c_3 + 2t^2 -2 c_0^2 x^2 \right) \\ & + c_2 \left(c_3^2 + 12 t^4 + 8 c_0^2 t^2 x^2 -4 c_0^4 x^4 + 8 c_3 (t^2 + c_0^2 x^2) \right)- 2 \Big(c_3^2 (-t^2 + c_0^2 x^2) \\ & -8 (t^2 - c_0^2 x^2)(t^2 + c_0^2 x^2)^2 + c_3 (-6 t^4 -4 c_0^2 t^2 x^2 + 2 c_0^4 x^4) \Big) \Bigg) \Bigg].
		\end{split}
	\end{equation}
	\begin{figure}[H]
		\centering
\subfigure[]{\includegraphics[width=0.37\linewidth]{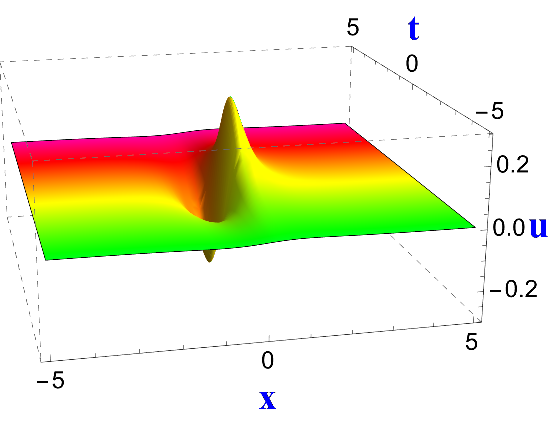}\label{lumps1-u-3d1}}	\subfigure[]{\includegraphics[width=0.36\linewidth]{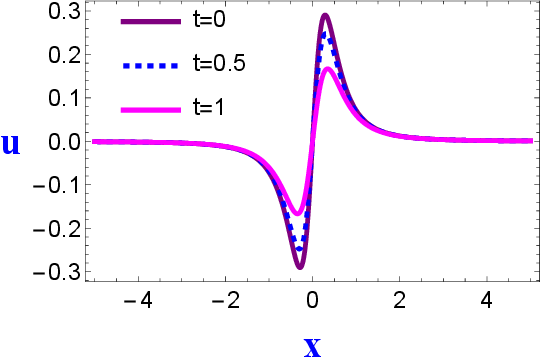}\label{lumps1-u-2d1}} 
		\subfigure[]{\includegraphics[width=0.25\linewidth]{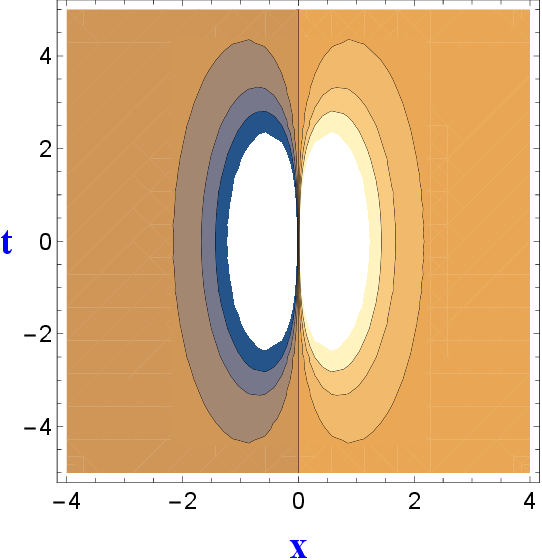}\label{lumps1-u-con1}} \\
		\subfigure[]{\includegraphics[width=0.33\linewidth]{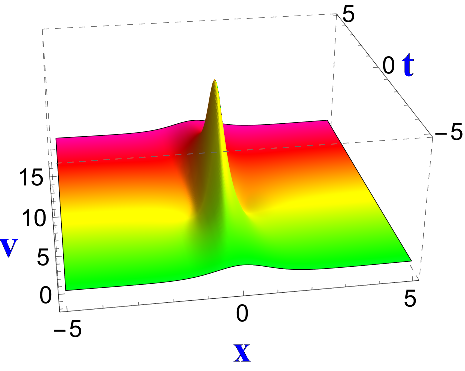}\label{lumps1-v-3d1}}
		\subfigure[]{\includegraphics[width=0.35\linewidth]{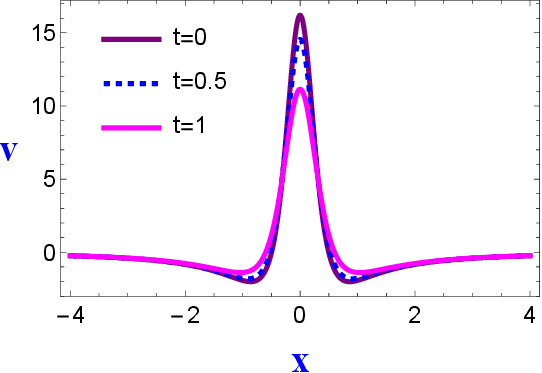}\label{lumps1-v-2d1}} 
		\subfigure[]{\includegraphics[width=0.26\linewidth]{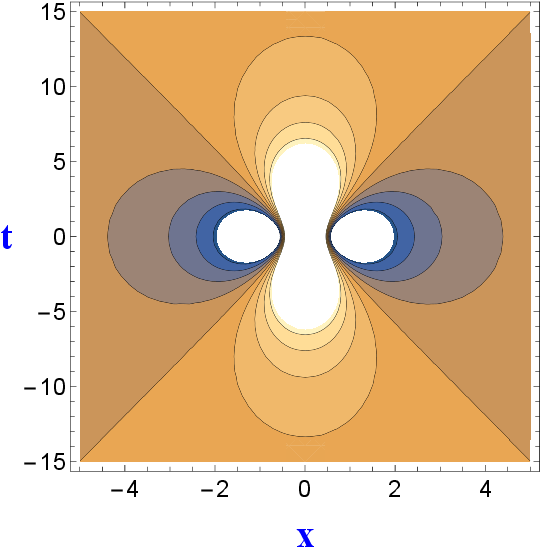}\label{lumps1-v-con1}} 
		\caption{Lump solutions for \eqref{lump2} when the parameters are $c_0=-3, c_2=4, c_3=5. $ (a), (d) are the 3D plots; (b), (e) are the wave propagation for different times; and (c), (f) are the corresponding contour plots for $u$ and $v$, respectively.}
		\label{fig13}
	\end{figure}
	The lump solutions $u$ and $v$ for \eqref{lump2} are plotted in Fig. \ref{fig13} by considering $c_0=-3, c_2=4, c_3=5$. From Figs. \ref{lumps1-u-3d1} and \ref{lumps1-v-3d1}, we observe that the lump waves have a localized characteristic at $t=0$. $v$ has two valleys symmetrically distributed over the peak (see Fig. \ref{lumps1-v-3d1}). The 3D plots confirm that lump solutions are algebraically decaying in the spatial direction. The propagation profiles in Figs. \ref{lumps1-u-2d1} and \ref{lumps1-v-2d1} clearly indicate that time has a negative impact on the amplitude of the lump solitons; as time increases, the amplitude of the peak gradually decreases. There is no visible movement for the lump solutions with respect to time in Figs. \ref{lumps1-u-2d1}, \ref{lumps1-v-2d1}. The solution $u$ has elliptic petal lump waves, as shown by Fig. \ref{lumps1-u-con1}. The solution $v$ also has petal lump waves; here the peak's trajectory is vertical, while the valley loops are horizontally distributed (see Fig. \ref{lumps1-v-con1}). The valley loops are bigger in Fig. \ref{lumps1-u-con1} than in Fig. \ref{lumps1-v-con1}.

	Various interesting structures can be obtained with a distinct set of values for $c_i$. For the system \eqref{m1}, a double lump structure appears for the parameter values $c_1=-\sqrt{-1} c_0, c_3=-c_2,$ where $c_0, c_2$ are free parameters. Now the double lump solutions for system \eqref{m1} can be expressed as 
	\begin{equation}\label{lump3}
		\begin{split}
			&u(x,t) =\frac{(4-4i)c_0 c_2 t}{c_2^2 -4t^2 \left(t+(1-i)c_0 x\right)^2}, \\
			& v(x,t)= \frac{16 i c_0^2 t^2 \Big(c_2^2 + 4t^2 \left(t+(1-i)c_0 x \right)^2 \Big)}{\Big(c_2^2 - 4t^2 \left(t+(1-i)c_0 x \right)^2 \Big)^2}.
		\end{split}
	\end{equation}
	
\begin{figure}[H]
		\centering
\subfigure[]{\includegraphics[width=0.23\linewidth]{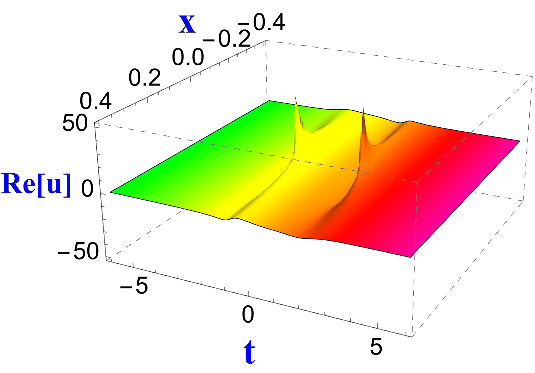}\label{dlump-u-re3d}} 
\subfigure[]{\includegraphics[width=0.23\linewidth]{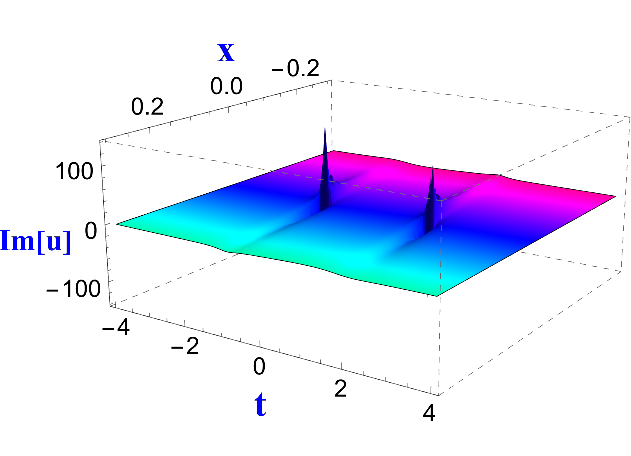}\label{dlump-u-im3d}}
\subfigure[]{\includegraphics[width=0.24\linewidth]{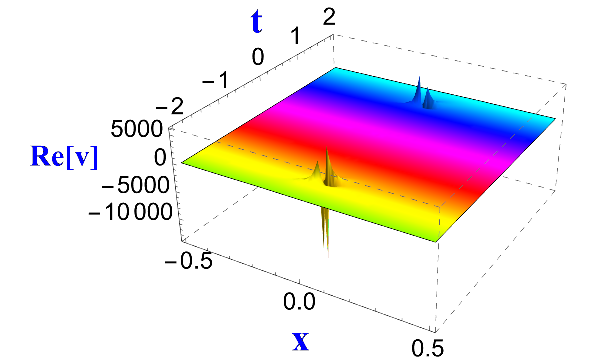}\label{dlump-v-re3d}} 
\subfigure[]{\includegraphics[width=0.28\linewidth]{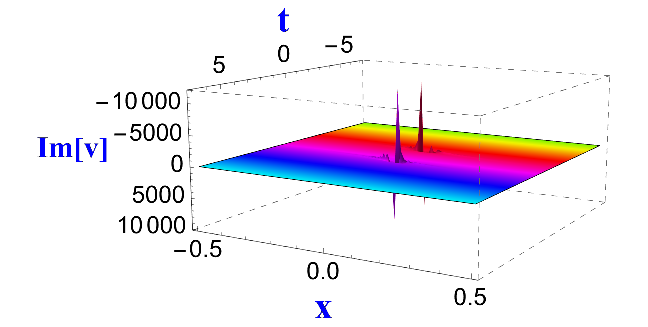}\label{dlump-v-im3d}} \\
\subfigure[]{\includegraphics[width=0.23\linewidth]{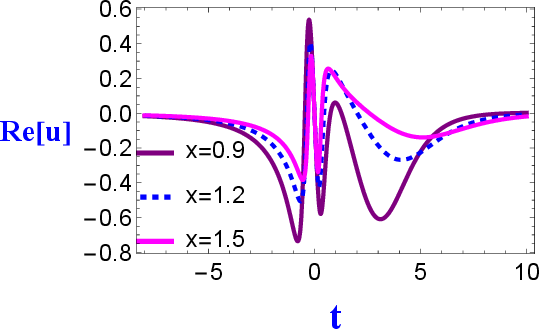}\label{dlump-u-re2d}} 
\subfigure[]{\includegraphics[width=0.23\linewidth]{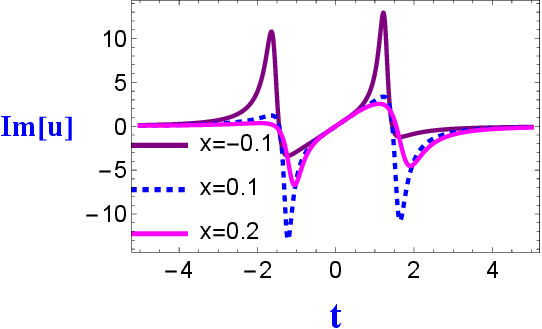}\label{dlump-u-im2d}} 
\subfigure[]{\includegraphics[width=0.23\linewidth]{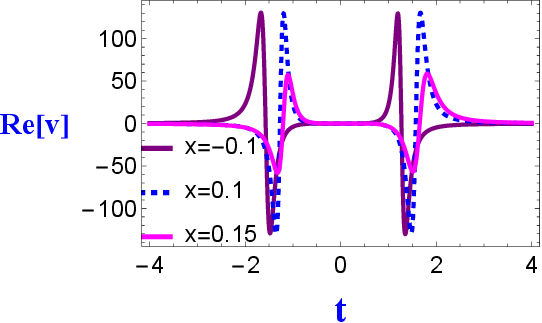}\label{dlump-v-re2d}} 
\subfigure[]{\includegraphics[width=0.23\linewidth]{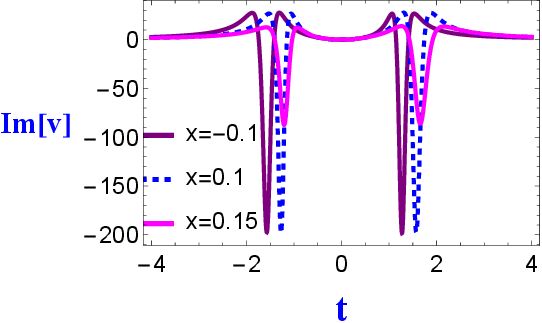}\label{dlump-v-im2d}} \\
		\subfigure[]{\includegraphics[width=0.23\linewidth]{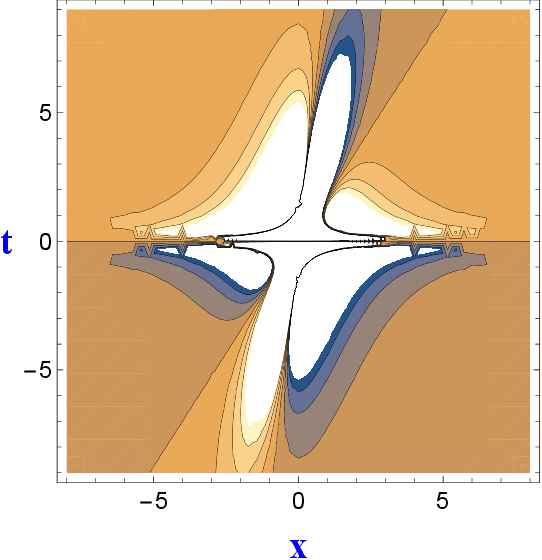}\label{dlump-u-recon}} 
		\subfigure[]{\includegraphics[width=0.23\linewidth]{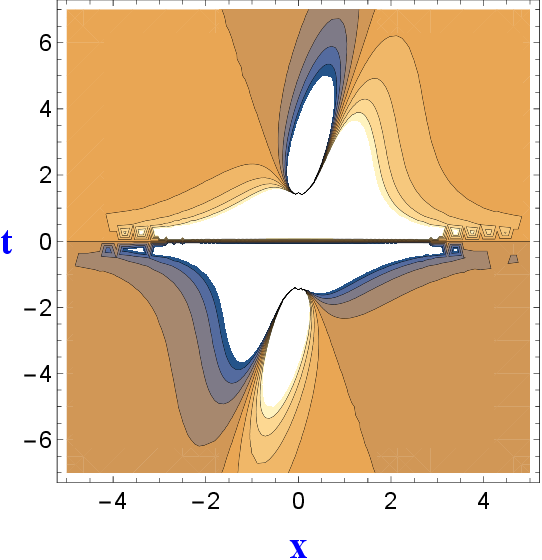}\label{dlump-u-imcon}}
		\subfigure[]{\includegraphics[width=0.23\linewidth]{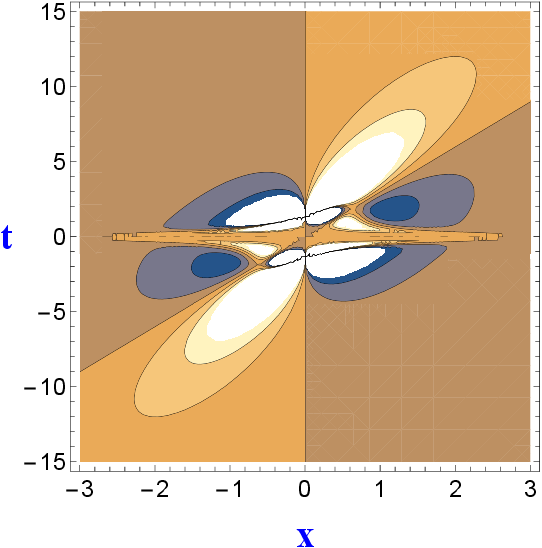}\label{dlump-v-recon}} 
		\subfigure[]{\includegraphics[width=0.23\linewidth]{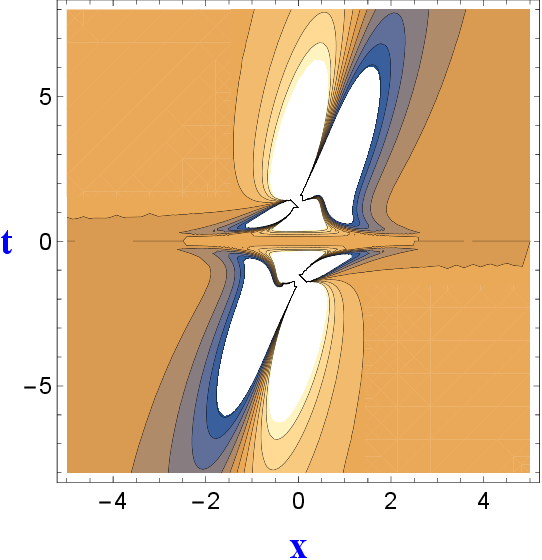}\label{dlump-v-imcon}}
		\caption{Double lump solutions for \eqref{lump3}. The 3D plots are represented by (a)-(d), the 2D plots by (e)-(h), and the contour plots (i)-(l) for the parameter set $c_0=-3, c_2=4.$ }
		\label{fig14}
	\end{figure}

	The existence of double lump solitons for the system \eqref{m1} is shown by Fig. \ref{fig14}, where the real and imaginary components of both solutions $u$ and $v$ are displayed. From the 3D plots in Figs. \ref{dlump-u-re3d}-\ref{dlump-v-im3d}, we observe that the double lumps appear for the time interval $-2 \leq t \leq 2$ in the neighborhood of $x=0$. They propagate with a constant background and decay to zero at infinity on the $(x,t)$-plane. This happened as lump waves are spatially localized. The 2D plots in Figs. \ref{dlump-u-re2d}-\ref{dlump-v-im2d} clearly visualize the occurrence of double lump solutions. As we shift the position towards the positive direction, the double lumps also moves along the positive spatial direction with decreasing amplitude. The contour plots in Figs. \ref{dlump-u-recon}-\ref{dlump-v-imcon} display a petal-like configuration for the double lump trajectories. The phase paths of two independent lumps are resemble about the horizontal axis, as shown in Figs. \ref{dlump-u-recon}-\ref{dlump-v-imcon}. 
	\section{Lump with one-kink}\label{lump1kink}
\begin{figure}[H]
		\centering
		\subfigure[]{\includegraphics[width=0.36\linewidth]{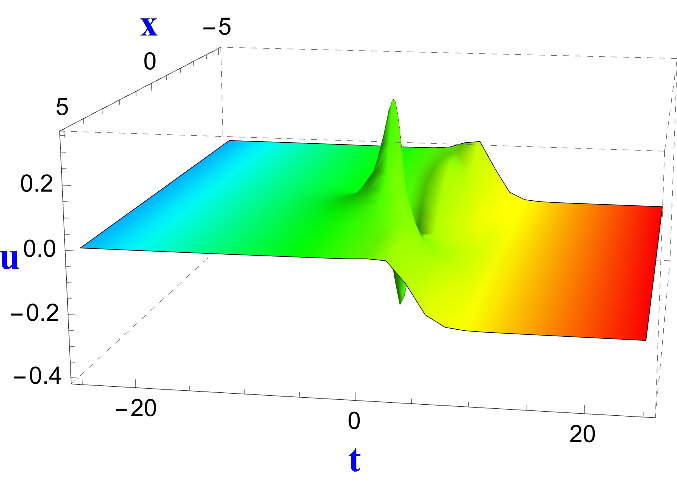}\label{lump1kinks1-u-3d1}} 
		\subfigure[]{\includegraphics[width=0.33\linewidth]{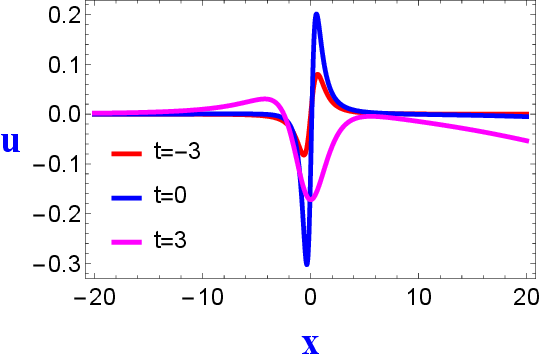}\label{lump1kinks1-u-2d1}}
		\subfigure[]{\includegraphics[width=0.24\linewidth]{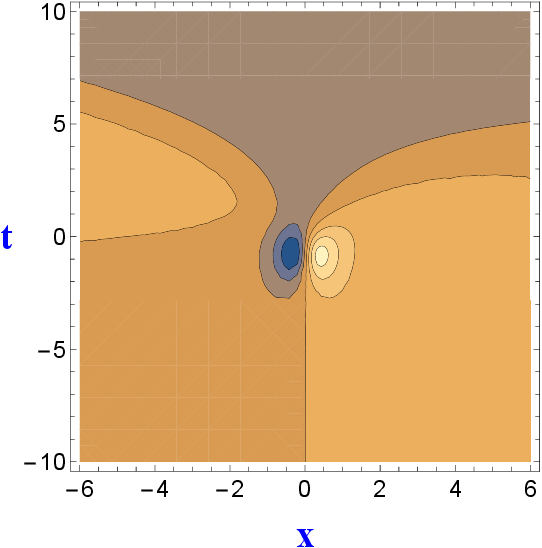}\label{lump1kinks1-u-con1}} \\
		\subfigure[]{\includegraphics[width=0.42\linewidth]{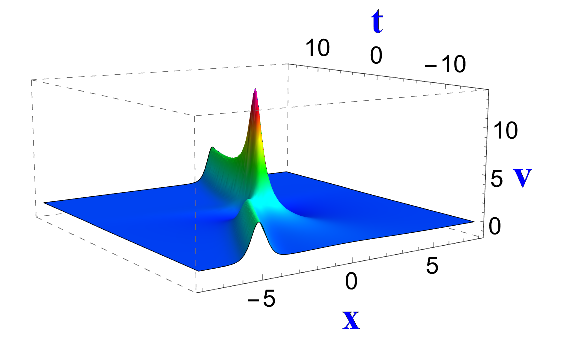}\label{lump1kinks1-v-3d1}}
		\subfigure[]{\includegraphics[width=0.31\linewidth]{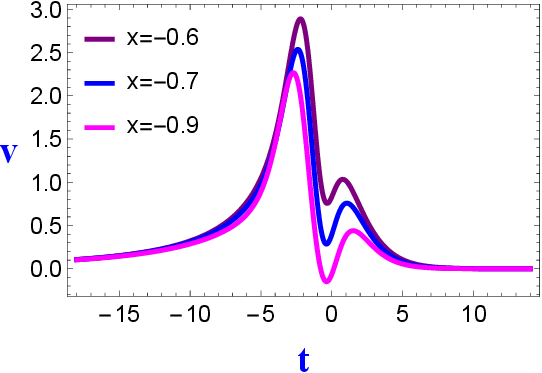}\label{lump1kinks1-v-2d1}} 
		\subfigure[]{\includegraphics[width=0.24\linewidth]{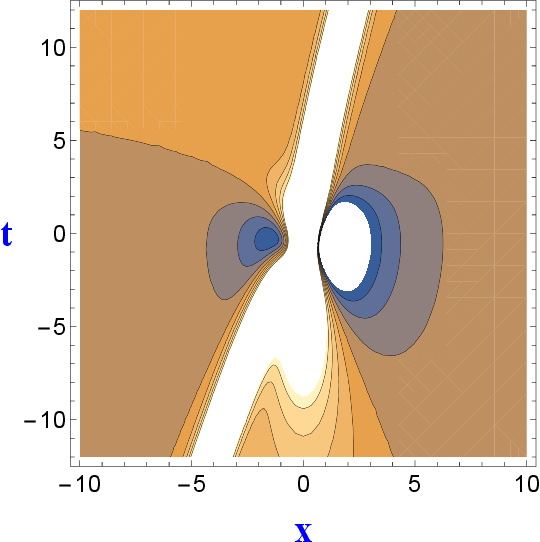}\label{lump1kinks1-v-con1}} 
		\caption{Lump with one kink for solution \eqref{lump1kinks1}, considering the first case of parameter set 1: $ c_0=-3, c_2=2, c_3=5, d_0=7.$ (a), (d) are the 3D plots, (b), (e) are the 2D plots, and (c), (f) are the corresponding contour plots for $u$ when $c_5=0.2,$ and $v$ when $c_5=-4$, respectively.}
		\label{fig15}
	\end{figure}

\begin{figure}[H]
		\centering
		\subfigure[]{\includegraphics[width=0.35\linewidth]{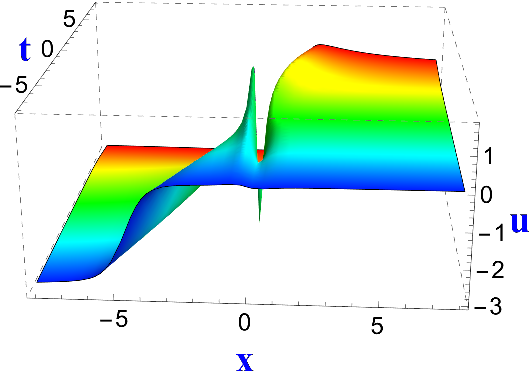}\label{lump1kinks2-u-3d1}} 
		\subfigure[]{\includegraphics[width=0.33\linewidth]{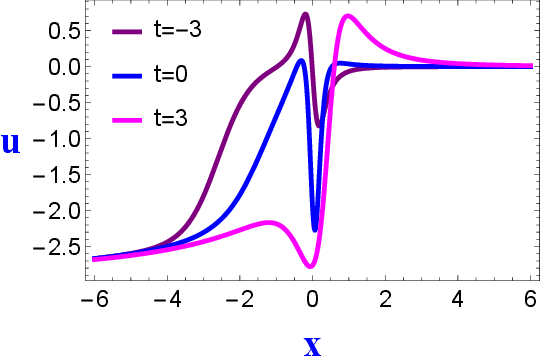}\label{lump1kinks2-u-2d1}}
		\subfigure[]{\includegraphics[width=0.26\linewidth]{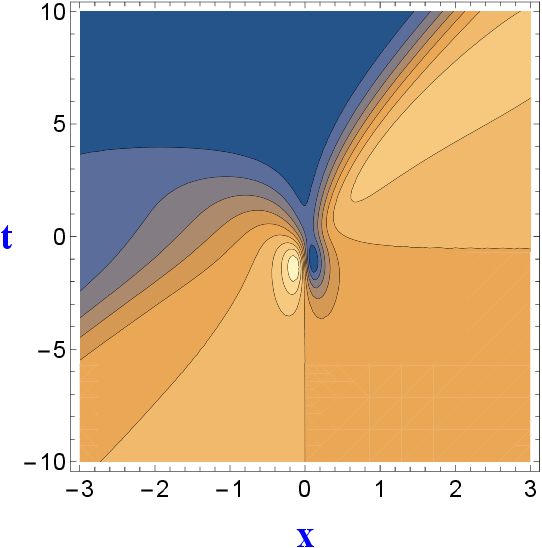}\label{lump1kinks2-u-con1}} \\
		\subfigure[]{\includegraphics[width=0.35\linewidth]{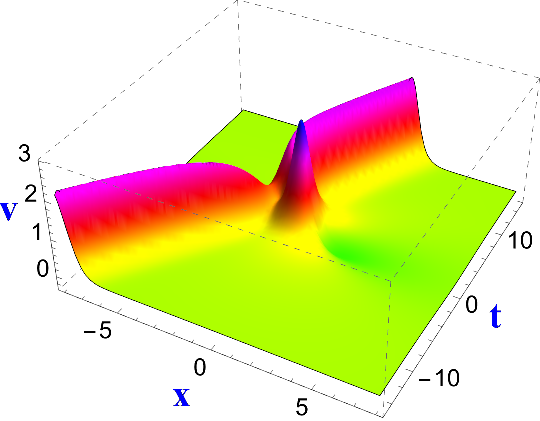}\label{lump1kinks2-v-3d1}}
		\subfigure[]{\includegraphics[width=0.33\linewidth]{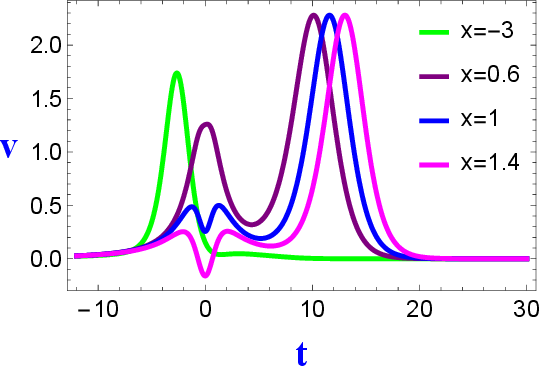}\label{lump1kinks2-v-2d1}} 
		\subfigure[]{\includegraphics[width=0.26\linewidth]{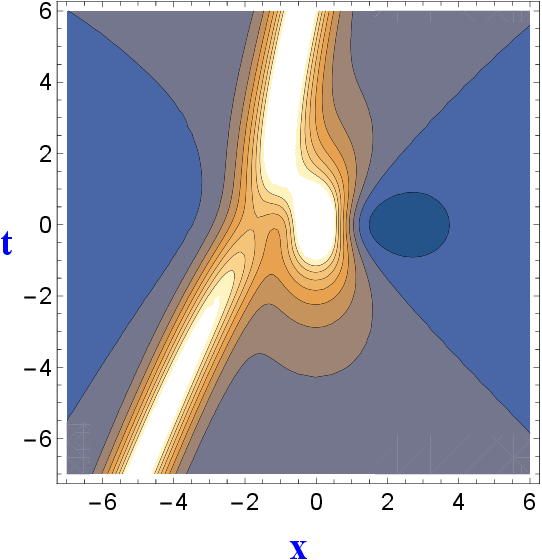}\label{lump1kinks2-v-con1}} 
		\caption{Lump with one-kink for solution \eqref{lump1kinks2}, considering the first case of set 2. (a), (d) are the 3D plots, (b), (e) are the 2D plots, and (c), (f) are the corresponding contour plots of $u$ when $c_0=-10, c_2=8, c_3=-1, c_4=-3, d_0=15$ and $v$ for $c_0=-1, c_2=2, c_3=5, c_4=-3, d_0=0.05$, respectively.}
		\label{fig16}
	\end{figure}

	Adding one exponential function \cite{nd2} to the positive quadratic expression \eqref{lump1}, one can determine the lump structure with the single kink wave. So, we consider the auxiliary functions $f(x,t)$ and $g(x,t)$ as follows  
	\begin{equation}\label{lump1kink1}
		f(x,t)=(c_0 x+t)^2 +(c_1 x+t)^2 +c_2+d_0 e^{c_4 x+t};~~ g(x,t)=(c_0 x+t)^2 +(c_1 x+t)^2 +c_3+d_0 e^{c_5 x+t}, 
	\end{equation}
	where $d_0$ and $c_i~(0 \leq i \leq 5)$ are the parameters to be determined. The following relations among the parameters are obtained by substituting the expressions \eqref{lump1kink1} into Eq. \eqref{quadratic}. \\
	\textbf{Set 1:} 
	\begin{equation*}
		\begin{cases}
			\begin{split}
				& c_1=-c_0, c_4=0; \\
				& c_0=\frac{1}{6} (-3 i \mp \sqrt{3}), c_1= \frac{1}{6} (-3i \pm \sqrt{3}), c_4=-i; \\ 
				& c_0=\frac{1}{6} (3 i \mp \sqrt{3}), c_1= \frac{1}{6} (3i \pm \sqrt{3}), c_4=i.
			\end{split}
		\end{cases}
	\end{equation*}
	For the first case of set 1, where $c_0, c_2, c_3, c_5,$ and $d_0$ are free parameters,
	the lump with one-kink of Eq. \eqref{m1} is given by 
	\begin{equation}\label{lump1kinks1}
		\begin{split}
			u(x,t) &= -\Bigg[\frac{-4 c_0^2 \Big(c_3 + d_0 e^t (-1+e^{c_5 x})\Big)x + c_2 \Big(c_5 d_0 e^{t+ c_5 x}+ 4 c_0^2 x \Big)+ c_5 d_0 e^{t+c_5 x} \Big(d_0 e^t + 2 (t^2+c_0^2 x^2)\Big)}{\Big(c_2 + d_0 e^t + 2(t^2+ c_0^2 x^2)\Big)\Big(c_3+d_0 e^{t+ c_5 x}+ 2(t^2+c_0^2 x^2)\Big)}\Bigg], \\
			v(x,t) &= \frac{1}{\Big(c_2 + d_0 e^t + 2 (t^2 + c_0^2 x^2)\Big)^2  \Big(c_3 + d_0 e^{t+c_5 x}+ 2(t^2 + c_0^2 x^2)\Big)^2} \Bigg[\Big(c_2 + d_0 e^t + 2 (t^2 + c_0^2 x^2)\Big) \\ &   \Big(c_3   +d_0 e^{t+c_5 x}+ 2(t^2+c_0^2 x^2)\Big) \Big(c_5^2 d_0 e^{t+ c_5 x} (c_2 + d_0 e^t + 2t^2)+ 48 c_0^4 x^2 + 2 c_0^2 (2c_2 + 2c_3 + 2 d_0 e^t  \\ & + 2 d_0 e^{t+c_5 x}+ 8t^2 + 4 c_5 d_0 e^{t+c_5 x}x + c_5^2 d_0 e^{t+c_5 x}x^2) \Big)- (c_5 d_0 e^{t+c_5 x}+ 4 c_0^2 x)\Big(c_2 + d_0 e^t + 2 (t^2 + c_0^2 x^2)\Big) \\ &  \Big(c_2 (c_5 d_0 e^{t+c_5 x}+ 4 c_0^2 x)+ c_5 d_0 e^{t+c_5 x}(d_0 e^t + 2 (t^2 + c_0^2 x^2))+ 4 c_0^2 x (c_3 + d_0 e^t (1+ e^{c_5 x})+ 4 (t^2 +c_0^2 x^2)) \Big) \\ &  -4 c_0^2 x(c_3 + d_0 e^{t+ c_5 x}+ 2(t^2+ c_0^2 x^2)) \Big(c_2 (c_5 d_0 e^{t+ c_5 x}+ 4 c_0^2 x)+c_5 d_0 e^{t+ c_5 x}(d_0 e^t + 2(t^2 + c_0^2 x^2)) \\ &  +4 c_0^2 x(c_3 + d_0 e^t(1+e^{c_5 x})+4 (t^2 + c_0^2 x^2)) \Big)  \Bigg].
		\end{split}
	\end{equation}
	Figs. \ref{fig15} and \ref{fig16} depict the lump wave along with one kink for  solutions \eqref{lump1kinks1} and \eqref{lump1kinks2}, respectively. A high-amplitude kink background is observed in Fig. \ref{lump1kinks1-u-3d1}, while the amplitude of the kink wave in Fig. \ref{lump1kinks1-v-3d1} is very low. In Fig. \ref{fig15}, the lump wave and the kink wave merge at the centre of the $(x,t)$-plane. From Fig. \ref{lump1kinks1-u-2d1}, we observe that the height of the peak of the lump wave is maximum at $t=0$. As we increase the time, the kink background gradually disappears, and as a result, the lump wave experiences a high impairment in amplitude. The kink shape was clearly visible after $x=0$, and beyond the time $t=1$, the kink trajectory began to bend and eventually disappeared; see the contour plot in Fig. \ref{lump1kinks1-u-con1}. As we shifted the position to the left, the peak's height gradually decreased (see Fig. \ref{lump1kinks1-v-2d1}). The left side valley went down completely due to the pressure exerted by the kink wave, whereas the right side valley formed a solitonic structure in the range $0\leq t\leq 5$, influenced by the kink wave (see Fig.\ref{lump1kinks1-v-2d1}). As a result, there is no loop for the left side valley in the contour plot Fig. \ref{lump1kinks1-v-con1}, but a remarkable phase shift is observed in Fig. \ref{lump1kinks1-v-con1}.
	The kink wave prevents the lump waves in Fig. \ref{fig16} from colliding on a line. Consequently, the lump wave is slightly shifted from the kink wave, but both are located in the neighborhood of the origin in the $(x,t)$-plane; see Figs. \ref{lump1kinks2-u-3d1}, \ref{lump1kinks2-v-3d1}. As time increases, the lump-kink wave takes on a perfect shape, as shown in Fig. \ref{lump1kinks2-u-2d1}. In reality, the solution $v$ fails to provide the satisfactory lump-kink wave in Fig. \ref{lump1kinks2-v-3d1}. Fig. \ref{lump1kinks2-v-2d1} justifies this statement. Initially, $v$ was a bright soliton, but as the position is shifted, it revealed a two-soliton structure at $ x = 0.6$. For further increment in the position, the smaller bright soliton was altered into a dark soliton structure at $x=1$ with minimum amplitude. Likewise, we observe that the amplitude of the newly revealed dark soliton gradually increases with the foregoing propagation of the taller bright soliton, as shown in Fig. \ref{lump1kinks2-v-2d1}. In these two-step variations, the amplitude of the taller, bright soliton is independent of position. The phase shift of Figs. \ref{lump1kinks2-u-3d1} and \ref{lump1kinks2-v-3d1} is shown by the contour plots Figs. \ref{lump1kinks2-u-con1} and \ref{lump1kinks2-v-con1}, respectively.
	\textbf{Set 2:} 
	\begin{equation*}
		\begin{cases}
			\begin{split}
				& c_1=-c_0, c_5=0; \\
				& c_0=\frac{1}{6} (-3  \mp i \sqrt{3}), c_1= \frac{1}{6} (-3 \pm i \sqrt{3}), c_5=-1; \\ 
				& c_0=\frac{1}{6} (3  \mp i \sqrt{3}), c_1= \frac{1}{6} (3 \pm i \sqrt{3}), c_5=1.
			\end{split}
		\end{cases}
	\end{equation*}
	Similarly, the first case of set 2, where $c_0, c_2, c_3, c_4,$ and $d_0$ are free parameters, yields another set of lump solutions with one kink for the system \eqref{m1}. 
	\begin{equation}\label{lump1kinks2}	
		\begin{split}
			u(x,t) &= \Bigg[\frac{-4 c_0^2 \Big(c_2 + d_0 e^t (-1+e^{c_4 x})\Big)x + c_3 \Big(c_4 d_0 e^{t+ c_4 x}+ 4 c_0^2 x \Big)+ c_4 d_0 e^{t+c_4 x} \Big(d_0 e^t + 2 (t^2+c_0^2 x^2)\Big)}{\Big(c_3 + d_0 e^t + 2(t^2+ c_0^2 x^2)\Big)\Big(c_2 +d_0 e^{t+ c_4 x}+ 2(t^2+c_0^2 x^2)\Big)}\Bigg], \\
			v(x,t) &= \frac{1}{\Big(c_3 + d_0 e^t + 2 (t^2 + c_0^2 x^2)\Big)^2  \Big(c_2 + d_0 e^{t+c_4 x}+ 2(t^2 + c_0^2 x^2)\Big)^2} \Bigg[\Big(c_3 + d_0 e^t + 2 (t^2 + c_0^2 x^2)\Big) \\ &   \Big(c_2   +d_0 e^{t+c_4 x}+ 2(t^2+c_0^2 x^2)\Big) \Big(c_4^2 d_0 e^{t+ c_4 x} (c_3 + d_0 e^t + 2t^2)+ 48 c_0^4 x^2 + 2 c_0^2 (2c_2 + 2c_3 + 2 d_0 e^t  \\ & + 2 d_0 e^{t+c_4 x}+ 8t^2 + 4 c_4 d_0 e^{t+c_4 x}x + c_4^2 d_0 e^{t+c_4 x}x^2) \Big)- (c_4 d_0 e^{t+c_4 x}+ 4 c_0^2 x)\Big(c_3 + d_0 e^t + 2 (t^2 + c_0^2 x^2)\Big) \\ &  \Big(c_3 (c_4 d_0 e^{t+c_4 x}+ 4 c_0^2 x)+ c_4 d_0 e^{t+c_4 x}(d_0 e^t + 2 (t^2 + c_0^2 x^2))+ 4 c_0^2 x (c_2 + d_0 e^t (1+ e^{c_4 x})+ 4 (t^2 +c_0^2 x^2)) \Big) \\ &  -4 c_0^2 x(c_2 + d_0 e^{t+ c_4 x}+ 2(t^2+ c_0^2 x^2)) \Big(c_3 (c_4 d_0 e^{t+ c_4 x}+ 4 c_0^2 x)+c_4 d_0 e^{t+ c_4 x}(d_0 e^t + 2(t^2 + c_0^2 x^2)) \\ &  +4 c_0^2 x(c_2 + d_0 e^t(1+e^{c_4 x})+4 (t^2 + c_0^2 x^2)) \Big)  \Bigg].
		\end{split}
	\end{equation}

	\section{Lump with two-kink}\label{lump2kinks}
	For this purpose we have to add two exponential functions \cite{lumpkink} to the expression \eqref{lump1}. To derive the lump solution with two-kink waves we assume the auxiliary functions $f(x,t)$ and $g(x,t)$ as follows 
	\begin{equation}\label{lump2kink}
		\begin{split}
			& f(x,t)=(k_1 x+k_2 t+k_3)^2 +(k_4 x+k_5 t+k_6)^2 +k_7+m_1  e^{b_1 x+b_2 t} + m_2 e^{b_3 x + b_4 t}, \\  & g(x,t)=(k_1 x+k_2 t+k_3)^2 +(k_4 x+k_5 t+k_6)^2 +k_8+m_1  e^{b_1 x+b_2 t} + m_2 e^{b_3 x + b_4 t},
		\end{split} 
	\end{equation}
	where $k_i~(1 \leq i \leq 8),$  $b_j~(j=1,2,3,4),$ and $m_1, m_2$ are the parameters to be measured. From Eqs. \eqref{quadratic} and \eqref{lump2kink}, a set of algebraic equations in independent variables can be obtained. The obtained equation set yields the following set of relations among the parameters
	\begin{equation}\label{lump2kink-exp}
		m_2=0, k_2 = \frac{2 k_3^2 k_4 k_5 -4 k_1 k_3 k_5 k_6 -2 k_4 k_5 k_6^2 + k_4 k_5 k_7 + k_4 k_5 k_8}{2 k_1 k_3^2 + 4 k_3 k_4 k_6 - 2 k_1 k_6^2 - k_1 k_7 -k_1 k_8},
	\end{equation}
	where $m_1, k_1, k_3, \dots, k_8, b_1, \dots, b_4 $ are free parameters. Hence considering the parameter set \eqref{lump2kink-exp},
	the expression of lump solutions with two-kink waves of Eq. \eqref{m1}  provided as
	\begin{equation}\label{lump2kink-sol}
		\begin{split}
			u(x,t)& = \frac{\Bigg[(-k_7 + k_8) \Bigg\{b_1 e^{b_1 x+ b_2 t} m_1 + 2 k_1 Q(x,t)  + 2 k_4 R(x,t) \Bigg\} \Bigg]} { \Bigg[   k_7 + e^{b_1 x+ b_2 t}m_1+ Q(x,t)^2  + R(x,t)^2  \Bigg] \Bigg[k_8 + e^{b_1x+b_2t}m_1+ Q(x,t)^2+R(x,t)^2  \Bigg]}, \\
			v(x,t) &= \frac{1}{\Big(k_7 + e^{b_1x+b_2t}m_1+ Q(x,t)^2 + R(x,t)^2 \Big)^2 \Big(k_8 + e^{b_1x+b_2t}m_1+ Q(x,t)^2 + R(x,t)^2 \Big)^2} \\ & \Bigg[  - \Big(b_1 e^{b_1 x+ b_2 t}m_1 + 2k_1 Q(x,t)+ 2 k_4 R(x,t) \Big)^2 \Big(k_7 + e^{b_1x+b_2t}m_1+ Q(x,t)^2 + R(x,t)^2 \Big)  \\&   \Big(k_7+ k_8 + 2e^{b_1x+b_2t}m_1+ 2Q(x,t)^2 + 2R(x,t)^2 \Big)    -  \Big(b_1 e^{b_1 x+ b_2 t}m_1 + 2k_1 Q(x,t)+ 2 k_4 R(x,t) \Big)^2  \\ &  \Big(k_8 + e^{b_1x+b_2t}m_1+ Q(x,t)^2 + R(x,t)^2 \Big)  \Big(k_7+ k_8 + 2e^{b_1x+b_2t}m_1+ 2Q(x,t)^2 + 2R(x,t)^2 \Big) \\& + \Big(k_7 + e^{b_1x+b_2t}m_1+ Q(x,t)^2 + R(x,t)^2 \Big)  \Big( k_8 + 2e^{b_1x+b_2t}m_1+ 2Q(x,t)^2 + 2R(x,t)^2 \Big) \\& \Big[2 \Big(b_1 e^{b_1 x+ b_2 t}m_1 + 2k_1 Q(x,t)+ 2 k_4 R(x,t) \Big)^2 + (2k_1^2 + 2k_4^2+ b_1^2 e^{b_1 x+ b_2 t}m_1) \\ & \Big(k_7 + e^{b_1x+b_2t}m_1+ Q(x,t)^2 + R(x,t)^2 \Big) +(2k_1^2 + 2k_4^2+ b_1^2 e^{b_1 x+ b_2 t}m_1) \\ & \Big(k_8 + e^{b_1x+b_2t}m_1+ Q(x,t)^2 + R(x,t)^2 \Big)  \Big] \Bigg],
		\end{split}
	\end{equation}
	where \[Q(x,t)=\Bigg(k_3 + \frac{k_5 (2 k_3^2 k_4-4 k_1 k_3 k_6+ k_4 (-2 k_6^2+ k_7 + k_8))t}{4 k_3 k_4 k_6+ k_1(2 k_3^2 -2 k_6^2-k_7 -k_8)} + k_1 x \Bigg),~~ 
	R(x,t)=(k_6 + k_5 t+ k_4 x). \]
	
\begin{figure}[H]
		\centering
		\subfigure[]{\includegraphics[width=0.35\linewidth]{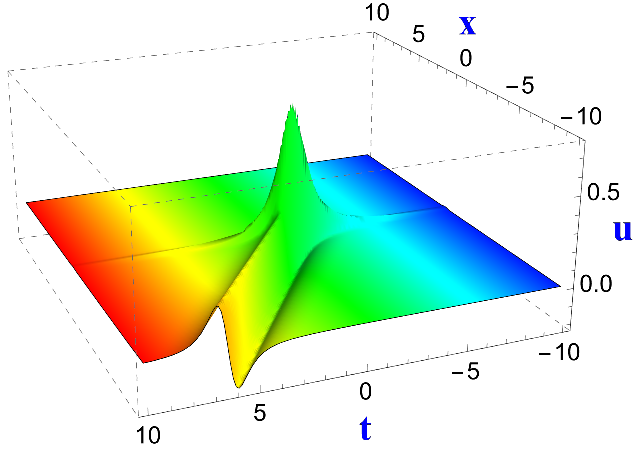}\label{lump2kink-u-3d1}} 
		\subfigure[]{\includegraphics[width=0.33\linewidth]{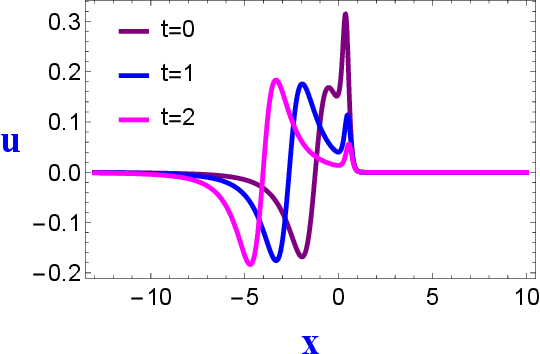}\label{lump2kink-u-2d1}}
		\subfigure[]{\includegraphics[width=0.26\linewidth]{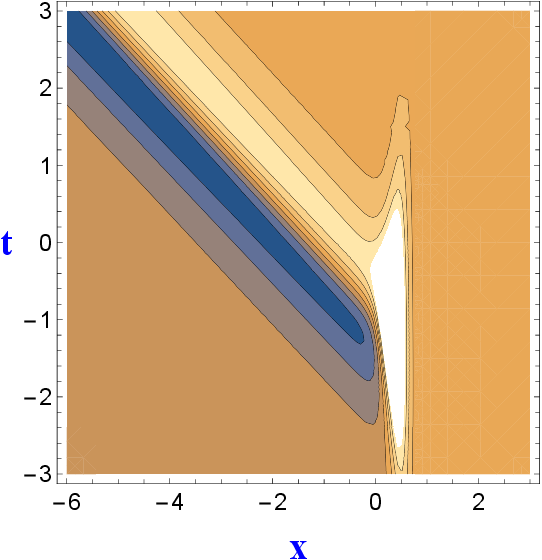}\label{lump2kink-u-con1}} \\
		\subfigure[]{\includegraphics[width=0.35\linewidth]{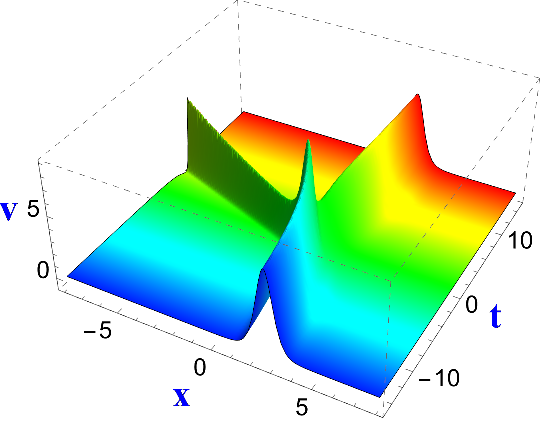}\label{lump2kink-v-3d1}}
		\subfigure[]{\includegraphics[width=0.33\linewidth]{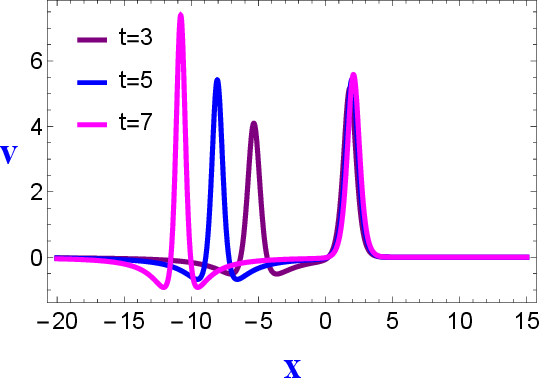}\label{lump2kink-v-2d1}} 
		\subfigure[]{\includegraphics[width=0.26\linewidth]{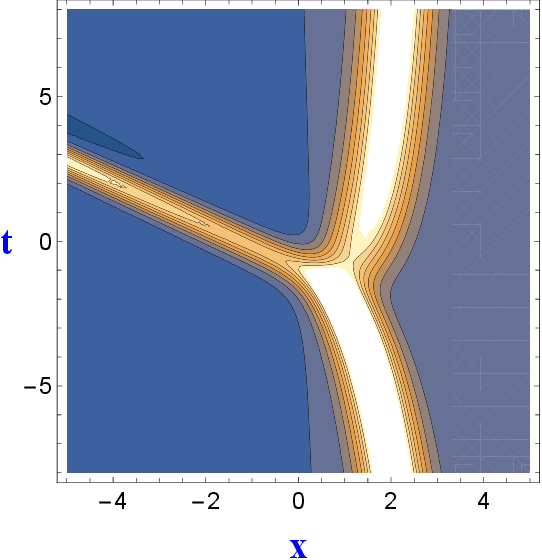}\label{lump2kink-v-con1}}\caption{Lump with two-kink for solution \eqref{lump2kink-sol}, when $ m_1=0.9, b_2=0.05, b_3=0.5, b_4=2, k_1=1.57, k_3=5.3, k_4=3, k_5=4.2, k_6=2.$ (a), (d) are the 3D plots, (b), (e) are the 2D plots, and (c), (f) are the corresponding contour plots of $u$ and $v$, respectively. For $b_1=10, k_7=0.5, k_8=6$, we have (a), (b), (c), and for $b_1=3.5, k_7=2.5, k_8=1.5$, we have (d), (e), (f).}
		\label{fig17}
	\end{figure}

	The lump solution with two-kink waves for \eqref{lump2kink-sol} is illustrated in Fig. \ref{fig17}. The dynamical behaviours are shown by the 3D plots in Figs. \ref{lump2kink-u-3d1}, \ref{lump2kink-v-3d1}. In Fig. \ref{lump2kink-u-3d1}, one kink wave appears obliquely with a large amplitude, while the other is horizontally spread over the $t$ axis. In Fig. \ref{lump2kink-u-2d1}, as time goes on, the lump soliton with two-kink waves moves along the negative spatial direction. After the position $x=2$, the effect of the kink wave on lump soliton vanishes, and it simply propagates as a plane wave in Fig. \ref{lump2kink-u-2d1}.The right side kink wave blocked the lump soliton from moving along the positive spatial direction in Fig. \ref{lump2kink-u-2d1}. At time $t=0$, the effect of the right-side kink wave was high, but it gradually decreased as time increased. Due to the impact of the left-side kink wave, the left side valley of the lump soliton disappears (see Fig. \ref{lump2kink-u-2d1}). The contour plot of Fig. \ref{lump2kink-u-3d1} is presented by the Fig. \ref{lump2kink-u-con1}. The solution $v$ for \eqref{lump2kink-sol} is plotted in Fig. \ref{lump2kink-v-3d1}. Here, the lump soliton acts as a knot of the two-kink waves. The propagation profile for $v$ shows a double-lump pattern due to the interaction between the lump soliton and two-kink waves, as shown in Fig. \ref{lump2kink-v-2d1}. There are two independent lump solitons in Fig. \ref{lump2kink-v-2d1}, and the effect of the kink wave is clearly visible on the right side lump, but the left side lump is entirely free from the kink wave. As time increases, the left side lump feels free to move along the negative spatial direction with a rapid increment in amplitude, while the movement of the right side lump is negligible along the positive spatial direction (see Fig. \ref{lump2kink-v-2d1}). Fig. \ref{lump2kink-v-con1} demonstrates the contour plot for the butterfly type structure of Fig. \ref{lump2kink-v-3d1}. Analytically, the solution $v$ for \eqref{lump2kink-sol} represents a lump wave with two- kinks, but in reality, $v$ shows a second-order lump structure in Fig. \ref{lump2kink-v-2d1}. This phenomenon occurs because the two kink waves, originating from opposite sides of the lump wave, interact with each other. As a result of this interaction, the two kink waves gain sufficient energy to occupy a higher amplitude, ultimately revealing a double lump pattern in Fig. \ref{lump2kink-v-2d1}. 
	
	\section{Interaction between breathers and other nonlinear waves}\label{bp} 
	Breather soliton is a localized, oscillatory solution to certain nonlinear PDEs. To understand the complex wave interactions in various physical systems, it is essential to study the breather structures. Breathers periodically change their amplitude or width, giving a pulsating appearance unlike the soliton without changing their shape while propagate. Breathers are distinct localized wave as lump solitons are fully localized in spatial dimensions but do not oscillate in the same way as breathers. In a wave packet, waves from various frequencies move at different speeds. So, they have a tendency to spread out over time, but the system's nonlinearity extinguishes the dispersion by altering the speeds of the various frequencies' waves inside the wave packet via interaction. Due to the balance between these two effects, the moving wave packet in a medium periodically oscillates between a compressed and stretched situation. The breathing of the wave packet reflects this oscillatory behaviour.
	Here, we will discuss the interaction of the first-order breather with various nonlinear waves. One can obtain a first-order breather by perturbing the auxiliary functions $f(x,t)$ and $g(x,t)$ up to second order \cite{mybreather} as follows
	\begin{equation}\label{bpi1}
		f(x,t)=1+ \epsilon f_1(x,t)+ \epsilon^2 f_2(x,t),~~
		g(x,t)= 1+ \epsilon g_1(x,t)+ \epsilon^2 g_2(x,t),
	\end{equation}
	where $ f_1(x,t)=a_1 e^{\theta_1}+ a_2 e^{\theta_2}, ~f_2(x,t)=a_{12} e^{\theta_1 + \theta_2}$ and $g_1(x,t)=b_1 e^{\theta_1} +b_2 e^{\theta_2}, ~g_2(x,t)= b_{12} e^{\theta_1 + \theta_2},$ with the phase $\theta_r=k_r x + \omega_r t,~ r=1,2,$ and $ b_{12}= \left(\frac{b_1 b_2}{a_1 a_2} \right) a_{12}, $
	\[	a_{12}= (a_1 a_2) \left(\frac{k_1 -k_2}{k_1 +k_2} \right) \Bigg[\frac{a_1 b_1 (a_2^2 -b_2^2)k_1^2 -a_2 b_2 (a_1^2 -b_1^2)k_2^2 -(a_1 +b_1)(a_2 + b_2)(a_2 b_1 -a_1 b_2)k_1 k_2}{a_1 b_1 (a_2^2 -b_2^2)k_1^2 + a_2 b_2 (a_1^2 -b_1^2)k_2^2 +(a_1 +b_1)(a_2 + b_2)(a_1 a_2  -b_1 b_2)k_1 k_2 } \Bigg].\] We have to restrict the wave number $k_r$ to a complex domain. So, we take $k_1=m+in,$ $k_2=m-in,~ m, n \in \mathbb{R}. $ Substituting the expression \eqref{bpi1} into Eq. \eqref{quadratic} we obtain the despersion relation $\omega_r= - \Big( \frac{a_r-b_r}{a_r+b_r}\Big) k_r^2, ~~r=1,2$. Therefore, $\theta_r,$ can be  expressed as $\theta_r= \Big[ \Big( mx- \left( \frac{a_r-b_r}{a_r + b_r} \right)(m^2 -n^2)t  \Big) \pm i \Big(n x- \left( \frac{a_r-b_r}{a_r + b_r}\right) 2mnt \Big) \Big], ~r=1,2. $
	Here $(+)$ and $(-)$ sign information of $i,$ refer for $\theta_1$ and $\theta_2$, respectively. After some calculations and setting $\epsilon=1$, the expression of the interacting breather for Eq.\eqref{m1} can be written as
	\begin{equation}\label{bpi2}
		\begin{split}
			& u(x,t)=\frac{\partial}{\partial x}\left[ln \left( \frac{f(x,t)}{g(x,t)} \right) \right],\;  v(x,t)= \frac{\partial^2}{\partial x^2}\Big[ln \left( f(x,t) g(x,t) \right)\Big], \text{where}  \\
			& f(x,t)=P(x,t)+i Q(x,t), \; g(x,t)=P'(x,t) + i  Q'(x,t),   \\ & P(x,t)= 1+a_1 e^A cosB + a_2 e^C cosD+ X e^{A+C} cos (B+D)-Y e^{A+C} sin(B+D),                     
			\\ & Q(x,t)= a_1 e^A sinB + a_2 e^C sinD+ X e^{A+C} sin (B+D)+ Y e^{A+C} cos(B+D),  \\ &
			P'(x,t)= 1+b_1 e^A cosB + b_2 e^C cosD+  \left( \frac{b_1 b_2}{a_1 a_2} \right) X e^{A+C} cos (B+D)- \left( \frac{b_1 b_2}{a_1 a_2} \right) Y e^{A+C} sin(B+D),   \\ & 
			Q'(x,t)= b_1 e^A sinB + b_2 e^C sinD+ \left( \frac{b_1 b_2}{a_1 a_2}\right) X e^{A+C} sin (B+D)+ \left( \frac{b_1 b_2}{a_1 a_2}\right) Y e^{A+C} cos(B+D), \\& A= \Big[ mx- \left( \frac{a_1-b_1}{a_1 + b_1} \right)(m^2 -n^2)t  \Big], \; B= \Big[ n x- \left( \frac{a_1-b_1}{a_1 + b_1}\right) 2mnt \Big] ,  \\ & C= \Big[ mx- \left( \frac{a_2-b_2}{a_2 + b_2} \right)(m^2 -n^2)t  \Big], \; D= \Big[ -n x + \left( \frac{a_2-b_2}{a_2 + b_2}\right) 2mnt \Big], 
			\\ & X= - a_1 a_2  \left( \frac{n}{m} \right) \frac{EG+FH}{G^2 +H^2},  \; Y= -a_1 a_2 \left( \frac{n}{m} \right) \frac{FG-EH}{G^2 +H^2} , 
			\\ & E= 2mn (a_1 b_2 + a_2 b_1) (a_1 a_2 -b_1 b_2), \; H= 2mn (a_2 b_1 - a_1 b_2) (a_1 a_2 + b_1 b_2),  \\ & F=(a_2 b_1 -a_1 b_2) \Big[(a_1 b_2 + a_2 b_1)m^2 + (a_1 b_2 + a_2 b_1 + 2 a_1 a_2 + 2 b_1 b_2)n^2  \Big] ,  \\ & G= (a_1 a_2 -b_1 b_2) \Big[(2 a_1 b_2 + 2 a_2 b_1+ a_1 a_2 + b_1 b_2)m^2 +(a_1 a_2 + b_1 b_2)n^2 \Big].
		\end{split}
	\end{equation}
	\begin{figure}[H]
		\centering
		\subfigure[]{\includegraphics[width=0.33\linewidth]{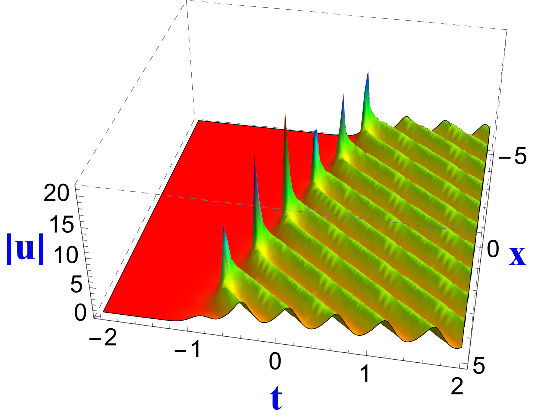}\label{bpi-u-3d1}} 
		\subfigure[]{\includegraphics[width=0.33\linewidth]{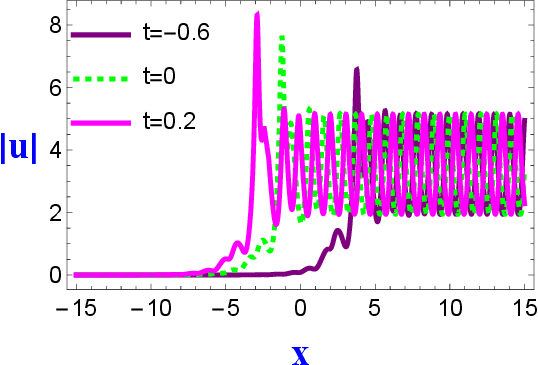}\label{bpi-u-2d1}}
		\subfigure[]{\includegraphics[width=0.24\linewidth]{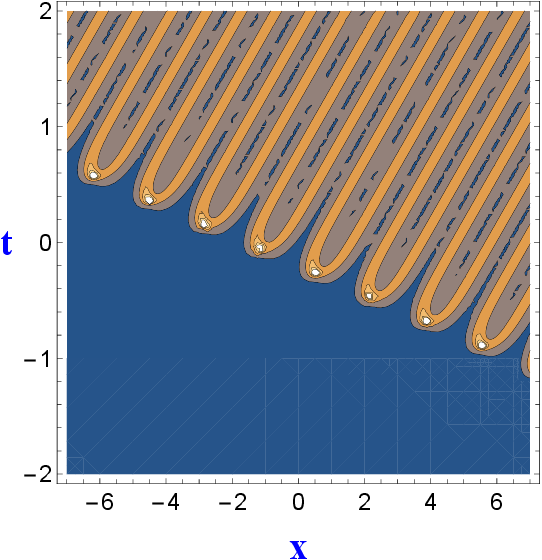}\label{bpi-u-con1}} \\
		\subfigure[]{\includegraphics[width=0.34\linewidth]{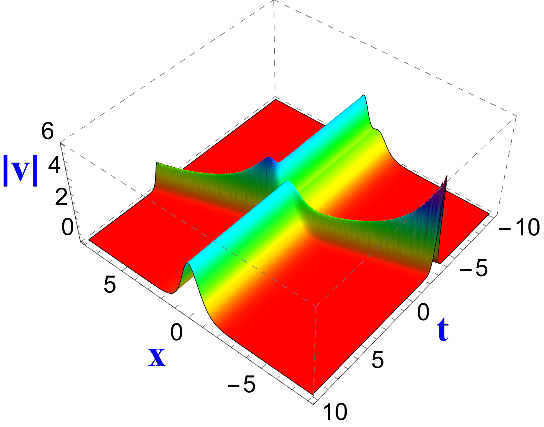}\label{bpi-v-3d1}}
		\subfigure[]{\includegraphics[width=0.33\linewidth]{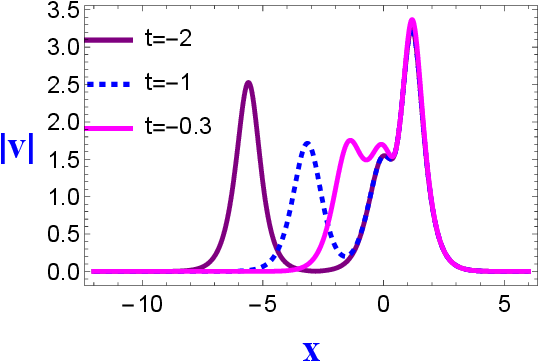}\label{bpi-v-2d1}} 
		\subfigure[]{\includegraphics[width=0.24\linewidth]{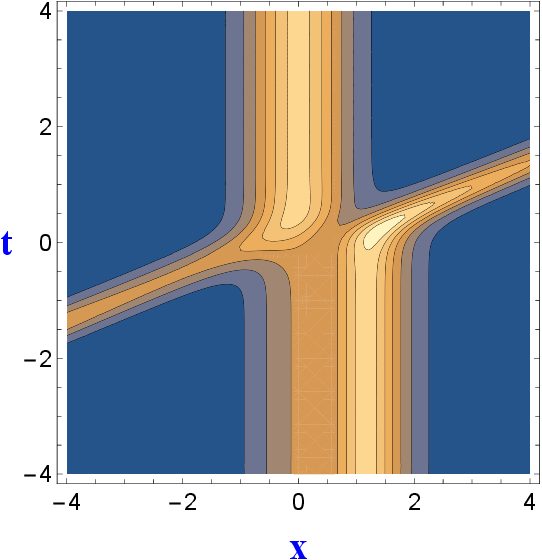}\label{bpi-v-con1}} 
		\caption{(a) Interaction between breathers and periodic waves; (d) interaction between the breather wave and bright soliton for solution \eqref{bpi2}. (a), (b), (c) appear for the parameters $a_1=1, a_2=4, b_1=0, b_2=0, m=1, n=3$, and (d), (e), (f) occur when $a_1=1, a_2=6, b_1=1, b_2=0, m=2.5, n=0.3$.}
		\label{fig18}
	\end{figure}
	For the parameter values $a_1=1, a_2=4, b_1=0, b_2=0, m=1, n=3$, the absolute value of the solution $u$ for \eqref{bpi2} depicts the interaction between the breathers and periodic waves, as shown in Fig. \ref{bpi-u-3d1}. Initially, $|u|$ was a plane wave; after interaction, the breathers release the periodic solutions, as shown by Fig. \ref{bpi-u-2d1}. As we know, a breather wave is a combination of periodic waves and solitons, so based on Fig. \ref{bpi-u-3d1}, we can say it is a fission-type interaction, where the breathing pattern disappears after interaction and leaves the periodic solutions alone. From Fig. \ref{bpi-u-2d1}, we can see that the breather waves confine in a certain region and don't vary over time, which proves that the breather is also a localized wave. As time increases, the periodic-breather waves propagate along the negative spatial direction with an increasing amplitude (see Fig. \ref{bpi-u-2d1}). The interacting phenomena between the breathers and periodic waves with a constant background are clearly depicted by the contour plot  Fig. \ref{bpi-u-con1}. 
	
	\begin{figure}[H]
		\centering
		\subfigure[]{\includegraphics[width=0.3\linewidth]{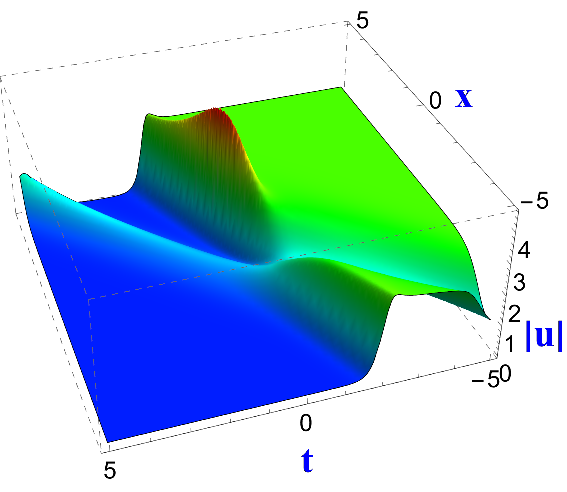}\label{bpi-u-3d3}} 
		\subfigure[]{\includegraphics[width=0.33\linewidth]{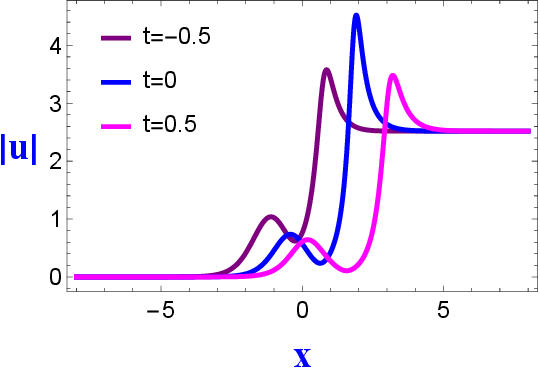}\label{bpi-u-2d3}}
		\subfigure[]{\includegraphics[width=0.24\linewidth]{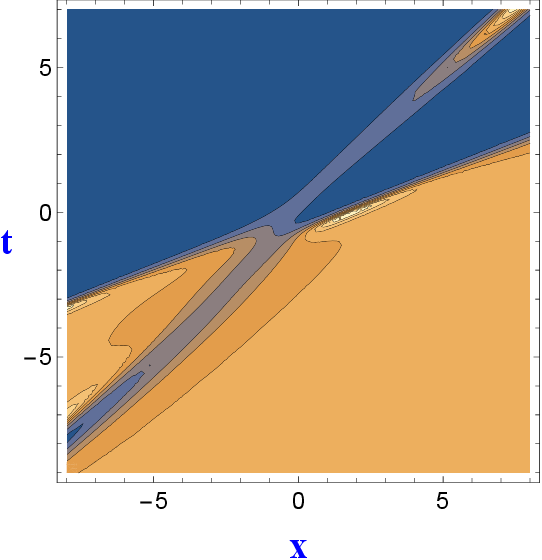}\label{bpi-u-con3}} \\
		\subfigure[]{\includegraphics[width=0.3\linewidth]{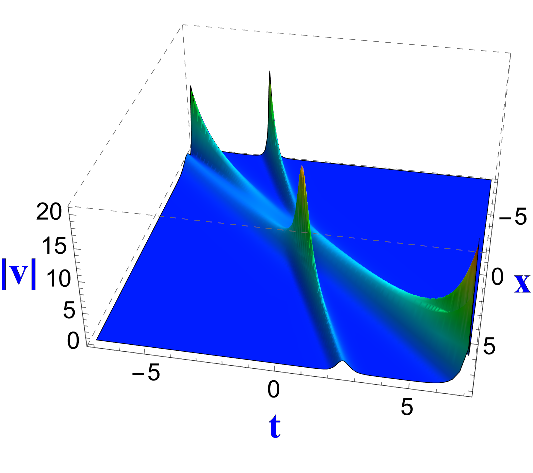}\label{bpi-v-3d3}}
		\subfigure[]{\includegraphics[width=0.35\linewidth]{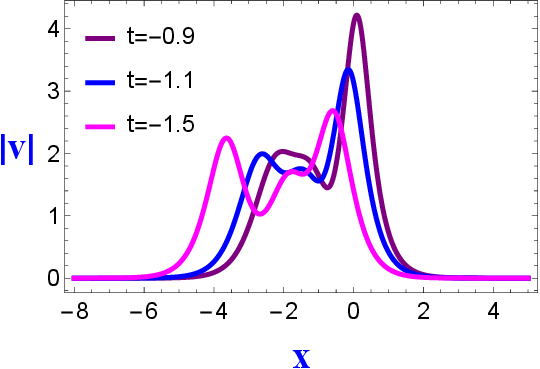}\label{bpi-v-2d3}} 
		\subfigure[]{\includegraphics[width=0.24\linewidth]{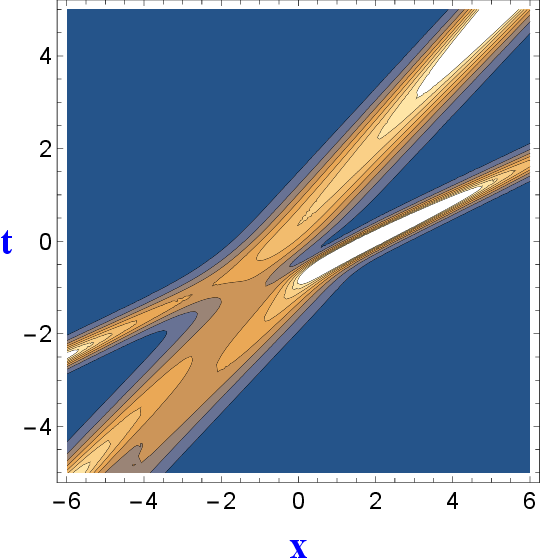}\label{bpi-v-con3}} 
		\caption{Considering the parameter values $a_1=1, a_2=4, b_1=0, b_2=1.5, m=2.5, n=0.3$, the interaction between the breather wave and the kink wave is shown in (a); the interaction between two harmony breathers is shown in (d).}
		\label{fig19}
	\end{figure}
	The solution $v$ for the parameter set $a_1=1, a_2=6, b_1=1, b_2=0, m=2.5, n=0.3$ describes the interaction between breather waves and bright soliton, as shown in Fig. \ref{bpi-v-3d1}. The interaction in Fig. \ref{bpi-v-3d1} occurs at the center of $(x,t)$ plane. Initially, the bright soliton separately propagates, but as time increases, it gains a faster velocity and completely merges with the breather waves, as shown in Fig. \ref{bpi-v-2d1}. The interaction distracts the path of the bright soliton into two tracks: one goes to the right and the other moves to the left with breathing excitation, as shown in the contour plot Fig. \ref{bpi-v-con1}.  \\
	The interaction between breather and kink wave is captured in Fig. \ref{bpi-u-3d3} for the absolute value of the solution $u$, when we consider $a_1=1, a_2=4, b_1=0, b_2=1.5, m=2.5, n=0.3$. The peak of the kinky-breather is maximum at $t=0$, but before and after the time $t=0$, it has the same amplitude and width, as shown in Fig. \ref{bpi-u-2d3}.
	Fig. \ref{bpi-u-con3} represents the contour plot of Fig. \ref{bpi-u-3d3}. Fig. \ref{bpi-u-3d3} shows how the breather structure and kink wave interact with each other for the parameter set $a_1=1, a_2=4, b_1=0, b_2=1.5, m=2.5, n=0.3$. For this same parameter set, in Fig. \ref{bpi-v-3d3}, we observe that two harmony breathers interact for the absolute component of the solution $v$. As time increases, the waves in Fig. \ref{bpi-v-2d3} move along the positive spatial direction, with an increment in amplitude. The result of Fig. \ref{bpi-v-3d3} is also described by the contour plot in Fig. \ref{bpi-v-con3}. 	
	\begin{figure}[H]
		\centering
		\subfigure[]{\includegraphics[width=0.35\linewidth]{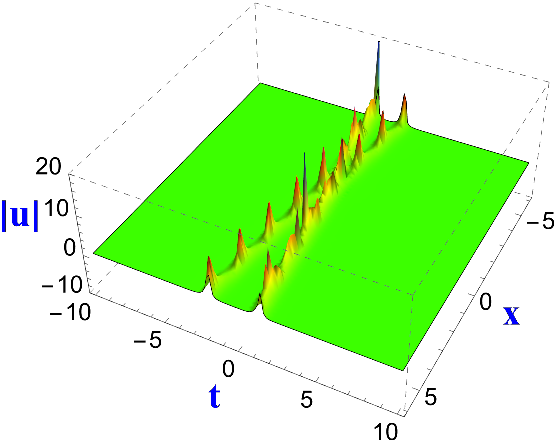}\label{bi-u-3d4}} 
		\subfigure[]{\includegraphics[width=0.33\linewidth]{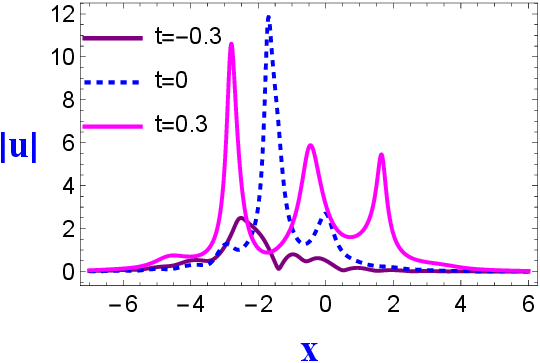}\label{bi-u-2d4}} 
		\subfigure[]{\includegraphics[width=0.27\linewidth]{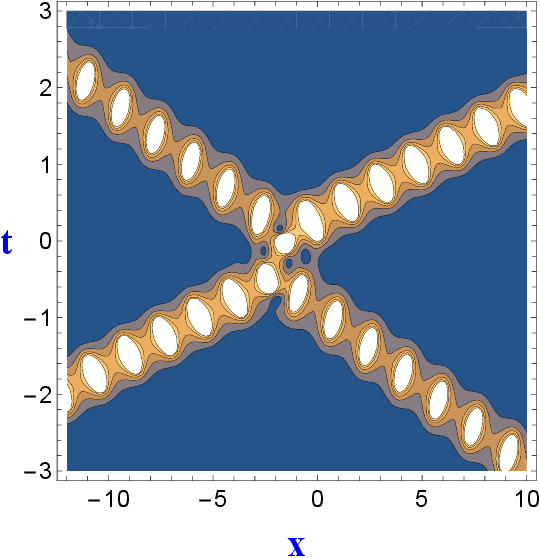}\label{bi-u-con4}} \\
		\subfigure[]{\includegraphics[width=0.35\linewidth]{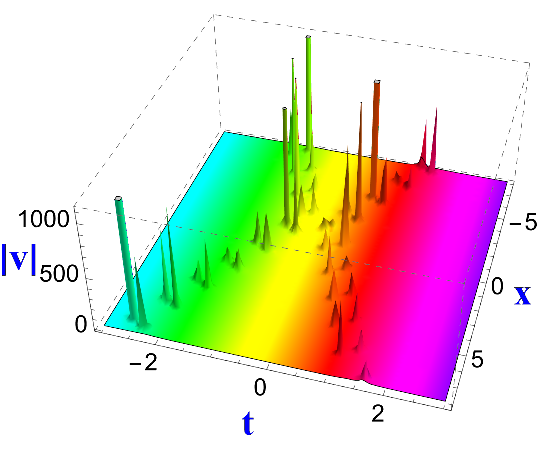}\label{bi-v-3d41}}
		\subfigure[]{\includegraphics[width=0.33\linewidth]{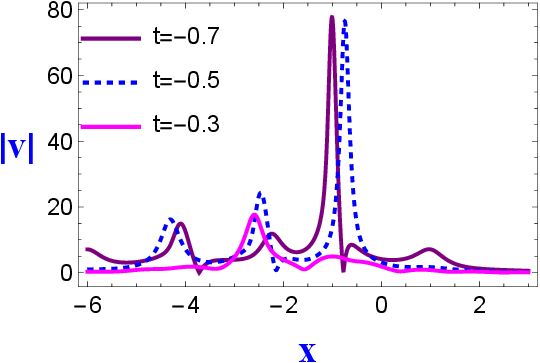}\label{bi-v-2d4}}
		\subfigure[]{\includegraphics[width=0.27\linewidth]{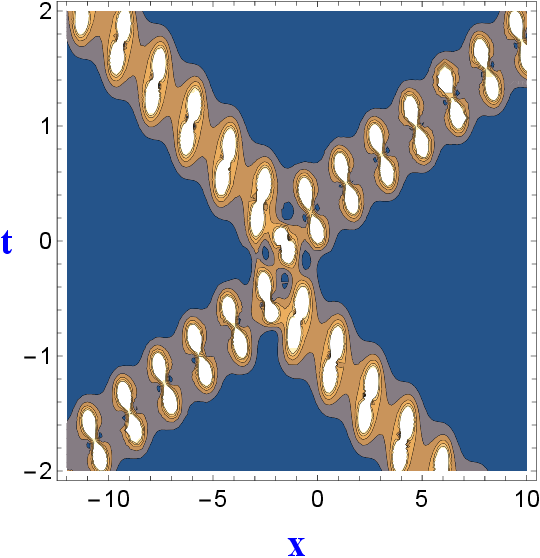}\label{bi-v-con4}} 
		\caption{(a) First-order breather-breather interaction; (d) second-order breather-breather interaction for solution \eqref{bpi2}, when $a_1=0.1, a_2=6, b_1=1.2, b_2=1.5, m=1, n=3$.  }
		\label{fig20}
	\end{figure}
	The breather-breather interaction is described by Fig. \ref{fig20} for the solution \eqref{bpi2}. For the parameter set $a_1=0.1, a_2=6, b_1=1.2, b_2=1.5, m=1, n=3$, $|u|$ exposes the interaction between two chains of first-order breathers (see in Fig. \ref{bi-u-3d4}), while $|v|$ discovers the interaction between two chains of second-order breathers (see in Fig. \ref{bi-v-3d41}). This type of breather wave can be traced from the negative space region, as shown in Fig. \ref{bi-u-2d4}, and gains its maximum peak at time $t=0$. Following the interaction, the wave assumes a larger amplitude compared to its pre-interaction state. The x-shape phase path in Fig. \ref{bi-u-con4} represents the contour of Fig. \ref{bi-u-3d4}. Fig. \ref{bi-u-con4} illustrates how, upon interaction, the left-sided breather chain shifts one unit more to the left and the right-hand breather chain shifts one unit more to the right. Two chains that simultaneously contain second-order breathers interact at the center of the space-time region, as shown by Fig. \ref{bi-v-3d41}. This type of breather is located in the negative space region; after the position $x>2$, it gradually disappears (see Fig. \ref{bi-v-2d4}). When we slightly increase the time, they reach near $x=0$, but if we increase the time more, the larger peak disappears, and the movement rapidly shifts towards the negative $x$-axis. This type of breather wave does not survive longer in the positive space region. The interaction between the chains of second-order breathers also represents an x-shape contour, as shown by Fig. \ref{bi-v-con4}. Fig. \ref{bi-v-con4} shows a slight change in the phase path after interaction. The widths of the breathers for $|u|$ is smaller than that of $|v|$ (see Figs. \ref{bi-u-3d4} and \ref{bi-v-3d41}); eventually, the intensity of Fig. \ref{bi-u-3d4} is greater than that of Fig. \ref{bi-v-3d41}. 

	\section{Conclusion} \label{con}   
	By employing the Hirota bilinear approach, we have successfully derived a diverse range of nonlinear waves for the CB system, including lump solitons, lump-kink wave interactions, breather-kink wave solutions, breather with bright soliton, and breather-breather solutions. The CB system is pivotal in various scientific and engineering applications, such as hydrodynamics, plasma physics, and optical physics, due to its rich dynamics and ability to model complex physical phenomena. Our results are presented graphically in 2D, 3D, and contour plots, providing visual insights into the behavior of these solitary waves. The Hirota bilinear form yields a family of lump solutions (see Figs. \ref{fig13}, \ref{fig14}), characterized by localized energy packets that remain stationary in all directions. By introducing exponential functions to a positive quadratic polynomial, we have demonstrated the interaction between lump soliton and kink wave (see Figs. \ref{fig15}–\ref{fig17}). Furthermore, our investigation shows that breather waves can interact with other nonlinear waves by incorporating complex conjugate wave numbers into the Hirota perturbation technique. The interaction between breather and periodic solution (see Fig. \ref{fig18}), breather-kink wave interactions (see Fig. \ref{fig19}), and breather-breather interactions (see Fig. \ref{fig20}) are illustrated in the provided figures. These graphical representations facilitate a deeper understanding of the dynamic behavior of the CB system and the interacting properties of the solitary waves. This method can be applied more broadly to handle several types of couple systems. The novel solutions obtained in this study may have important applications in recognizing and investigating various physical phenomena in fields such as laser optics, communication systems, fluid dynamics, heat transfer, shallow water waves, acoustics, and optics.\\\\
	
	\footnotesize{	\textbf{Acknowledgments}
		Snehalata Nasipuri is thankful to the Council of Scientific and Industrial Research (CSIR), India, for providing financial support under the Senior Research Fellowship (SRF) program (File No: 09/202(0120)/2021-EMR-I).
		\\\\
		\textbf{Author contributions}
		SN contributed to writing original draft, methodology, software, and visualization. PC contributed to investigation, conceptualization, and supervision. 
		\\\\
		\textbf{Funding} No funding was received to assist with the preparation of this manuscript. Also, the author has no financial or proprietary interests in any material discussed in this article. \\\\
		\textbf{Data availability statement}
		The author confirms that there is no associated data available for the above research work. Data sharing does not apply to this article as no new data were created or analyzed in this study.  \\\\
		\textbf{Declarations} \\\\
		\textbf{Conflict of interest}
		The authors have no conflict of interest.}

\end{document}